\documentclass[useAMs,usenatbib,a4paper,fleqn]{mn2e}
\usepackage{savesym}
\usepackage{graphicx}
\expandafter\let\csname equation*\endcsname\relax
  \expandafter\let\csname endequation*\endcsname\relax 
\usepackage{subfig}
\usepackage{amsmath}
\usepackage{amssymb}
\usepackage{verbatim}
\usepackage{array}
\usepackage{times}
\usepackage[total={17.8cm,24.0cm},centering]{geometry} 

\newcommand{\be}{\begin{equation}}
\newcommand{\beq}{\begin{equation}}
\newcommand{\ee}{\end{equation}}
\newcommand{\eeq}{\end{equation}}
\newcommand{\eea}{\end{eqnarray}}
\newcommand{\bea}{\begin{eqnarray}}

\newcommand\bb[1] { \mbox{\boldmath{$#1$}} }

\def\gtsima{$\; \buildrel > \over \sim \;$}
\def\gtsim{\lower.5ex\hbox{\gtsima}}

\def\ltsima{$\; \buildrel < \over \sim \;$}
\def\ltsim{\lower.5ex\hbox{\ltsima}}

\title[The slow decline of T Pyx, IM Nor, and CI Aql]{The slow decline of the Galactic recurrent novae T Pyxidis, IM Normae, and CI Aquilae}
\author[Andrea Caleo and Steven N. Shore]{Andrea Caleo$^1$\thanks{E-mail: andrea.caleo@astro.ox.ac.uk} and Steven N. Shore$^{2,3}$
\\
$^1$Oxford Astrophysics, Denys Wilkinson Building, Keble Road, Oxford, OX1 3RH, United Kingdom\\
$^2$Dipartimento di Fisica ``Enrico Fermi'', University of Pisa, 56127 Pisa, Italy\\
$^3$INFN - Sezione di Pisa, largo B. Pontecorvo 3, 56127 Pisa, Italy}
\begin{document}
\date{}

\pagerange{\pageref{firstpage}--\pageref{lastpage}} \pubyear{2014}

\maketitle

\label{firstpage}

\begin{abstract}
A distinguishing trait of the three known Galactic recurrent novae with the shortest orbital periods, T Pyx, IM Nor, and CI Aql, is that their optical decline time-scales are significantly longer than those of the other recurrent systems. On the other hand, some estimates of the mass of the ejecta, the velocity of the ejecta, and the duration of the soft X-rays emission of these systems are of the order of those of the other recurrent systems and the fast classical novae. \\
We put forth a tentative explanation of this phenomenon. We propose that in these systems part of the material transferred from the companion during the first few days of the eruption remains within the Roche lobe of the white dwarf, preventing the radiation from ionizing the ejecta of the system and increasing the optical decline time-scale. We explain why this phenomenon is more likely in systems with a high mass transfer rate and a short orbital period. Finally, we present a schematic model that shows that the material transferred from the companion is sufficient to absorb the radiation from the white dwarf in these systems, ultimately supporting this scenario as quantitatively realistic. 
\end{abstract}

\maketitle

\section{Introduction}
Classical novae are the result of a thermonuclear runaway induced in the envelope of a mass-accreting white dwarf. The white dwarf is not destroyed by the process and the system may undergo subsequent eruptions. Recurrent novae, the rarest subclass, are those systems that have been observed to experience more than one historical eruption, with a recurrence time-scale of a few decades. These are characterised by a high mass transfer rate as well as a high white dwarf mass (see e.g. \citealt{Anupama2008}). Only ten Galactic recurrent novae have been directly observed during at least two of their eruptions; the statistics of these observations are reported by \citet{Schaefer2010} and are partially summarised here in table \ref{table:RN}. 

\begin{table*}
\centering
\small
\caption{The Galactic recurrent novae from \citet{Schaefer2010} with the addition of the latest eruptions of U Sco (2010) and T Pyx (2011). The systems are listed in order of orbital period. We have not included the eruption of T Pyx close to 1866, inferred by \citet{SchaeferPagnotta2010}.}
\begin{tabular}{c c c c c p{7 cm}}
\hline
\hline 
RN & $V_{\text{peak}}$ (mag) & $V_{\text{min}}$ (mag) & $t_3$ (d) & $P_{\text{orb}}$ (d) & Eruption years \\[0.5ex]
\hline
T Pyx & 6.4 & 15.5 & 62& 0.076&1890, 1902, 1920, 1944, 1967, 2011\\
IM Nor & 8.5 & 18.3 &  80& 0.102&1920, 2002\\
CI Aql & 9.0 & 16.7 & 32& 0.62&1917, 1941, 2000\\
V2487 Oph & 9.5 & 17.3 & 8&  $\sim$ 1&1900, 1998\\
U Sco & 7.5 & 17.6 & 2.6& 1.23& 1863, 1906, 1917, 1936, 1945, 1969, 1979, 1987, 1999, 2010\\
V394 CrA & 7.2 & 18.4 & 5.2& 1.52& 1949, 1987\\
T CrB & 2.5 & 9.8 &6 & 228& 1866, 1946 \\
RS Oph & 4.8 & 11 & 14& 457& 1898, 1907, 1933, 1945, 1958, 1967, 1985, 2006\\
V745 Sco & 9.4 & 18.6 & 9& 510& 1937, 1989\\
V3890 Sgr & 8.1 & 15.5 & 14& 519.7& 1962, 1990\\
\hline
\end{tabular}
\label{table:RN}
\end{table*}

In this paper we highlight an interesting feature of the data in table \ref{table:RN}: the three recurrent novae with the shortest orbital periods T Pyx, IM Nor, and CI Aql have decline time-scales $t_3$ significantly longer than all the other systems. In section \ref{sec:slowdecline} we argue that there is some evidence for the other observed properties of these systems to be in line with the other recurrent novae. We discuss the mass of their ejecta in sections \ref{sec:massejecta} and \ref{sec:massejecta2}, and the duration of their soft X-rays emission phases and the velocity of the ejecta in section \ref{sec:xrays}. In section \ref{sec:effect} we put forward a tentative explanation for this phenomenon. Finally, in section \ref{sec:conclusion} we summarize our findings.

\section{T Pyx, IM Nor, CI Aql, and the other novae} \label{sec:slowdecline}

The latest outburst of the recurrent nova T Pyx occurred in 2011 and the development of its light curve and spectrum has been followed more thoroughly than for any other recurrent novae to date (e.g. see \citealt{ShoreETAL2012} and \citealt{ShoreETAL2013} for the optical and UV spectra, \citealt{DeGennaroETAL2014} and \citealt{Chomiuk2014} for the X-rays, \citealt{Nelson2014} for the radio, and \citealt{Patterson2014} for the time-evolution of the orbital period). This allows astronomers to compare the features of T Pyx with those of other classical and recurrent novae.

\subsection{The mass of the ejecta} \label{sec:massejecta}
It is difficult to accurately determine the mass of the ejecta of a nova. It has usually been estimated based on photoionization models alone, which requires panchromatic observations (at least including the ultraviolet and optical resonance and principal recombination lines).  Recently, a new technique has been applied to several bright novae including T Pyx, to obtain density, filling factor, and mass estimates based on electron density determinations using forbidden lines from the nebular stage spectra (see \citealt{ShoreETAL2012} and \citealt{ShoreETAL2013}). In brief, either or both of the isoelectronic forbidden lines of [N II] and [O III] provide determinations of the density and electron temperature.  The electron density map is obtained by line profile ratios instead of using integrated fluxes assuming that the nebular spectra are (roughly) isothermal. The line profiles are modeled adopting a bipolar structure with a linear velocity law \citep{Ribeiro2013} to obtain the geometry of the isothermal ejecta. Knowledge of the distance from the object allows us to convert the H$\beta$ and H$\alpha$ fluxes to emission measures that are then used to obtain the filling factor using the independently obtained electron densities and volumes. This is then used to obtain the total mass of the ejecta. The temporal development of the electron density can also be determined using multi-epoch modeling \citep{Schwarz2014}.

Studies have been conducted for CI Aql by \citet{Iijima2012} and for T Pyx by \citet{ShoreETAL2013}, resulting in the value $M_{\text{ej}} \approx 2 \cdot 10^{-6} $ M$_\odot$ for both systems. This number is on the lower end of the scale for ejecta masses of classical novae. \citet{ShoreETAL2013} also quote a value for the filling factor of T Pyx, $f \approx 3 \cdot 10^{-2}$. To our knowledge, there is no estimate of the mass of the ejecta of IM Nor.  

We note that the value of the mass of the ejecta of T Pyx is controversial. While the spectroscopic value is low, there are indications that support a higher value, the most suggestive of which is based on the orbital period change undergone by the system during the eruption \citep{Patterson2014} and gives a mass of the ejecta $M_{\text{ej}} \geq 3 \cdot 10^{-5}$ M$_{\odot}$. We argue that the dynamics of the binary system during the eruption is complex and would deserve further consideration, discussing this issue in appendix A.

Another indication of a possible high value for the mass of the ejecta of T Pyx is provided by radio observations \citep{Nelson2014}. The peculiarity of T Pyx as a nova is evident in this spectral range, as the radio flux began rising surprisingly late ($\sim$ 50 days after the outburst), and no simple model of instantaneous, homologous explosion is able to satisfyingly fit the data and provide an estimate for $M_{\text{ej}}$. However, a model designed to fit to the later part of the light curve gives $M_{\text{ej}} \approx 4 \cdot 10^{-4}$ M$_\odot$, and more elaborate models for the whole light curve suggest a range of $(1 - 30) \cdot 10^{-5}$ M$_\odot$.

\subsection{$M_{\text{ej}} - t_2$ correlation} \label{sec:massejecta2}
\begin{figure}
\begin{center}
\includegraphics[width=0.5\textwidth]{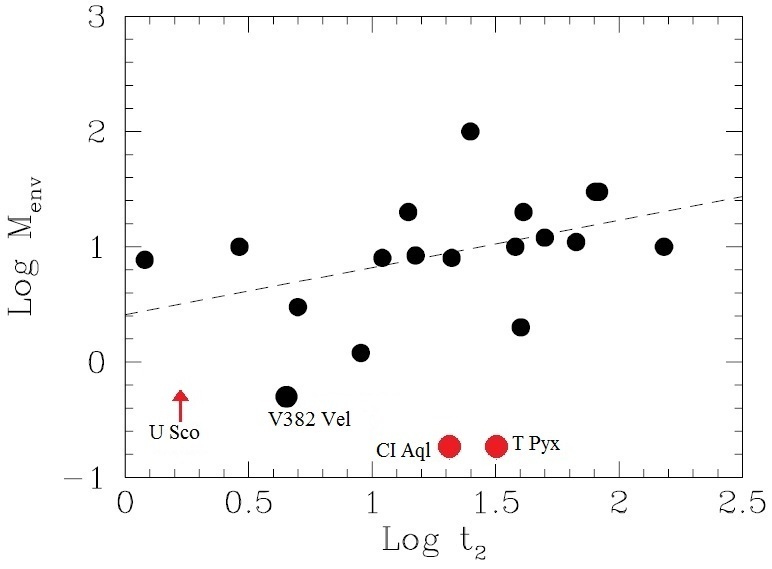}
\caption{\label{figEjectat2}Optical decline time-scale $t_2$ and mass of the ejecta for a set of novae; $M_{\text{ej}}$ is measured in units of $10^{-5} $ M$_\odot$ and $t_2$ in days. The approximate positions of the recurrent systems T Pyx, CI Aql, and U Sco, based on estimates of $t_2$ from the light curve of T Pyx, on the light curves reported by \citet{Schaefer2010} and on the masses of the ejecta estimated by \citet{ShoreETAL2013}, \citet{Iijima2012} and \citet{DiazETAL2012} ($M_{\text{ej}} > 3 \cdot 10^{-6}$ M$_\odot$ for U Sco) have been marked in red. V382 Vel is highlighted in the article by \citet{DellaValleETAL2002} because that system was one of the main objects of study of their article. The dashed curve is given by equation \eqref{C2eqCorrel} and it is the best linear fit for the $Log(M_{\text{ej}})\ -\ Log(t_2)$ relation.}
\end{center}
\end{figure}

Low-mass ejecta have lower density than high-mass ones and they are also expanding at higher speed so that their density will drop faster. The ionization of the ejecta will therefore be fast if the mass of the ejecta is low, and the time for the decline will be accordingly short, so that a correlation between these quantities can be expected. Despite the uncertainties in the masses, \citet{DellaValleETAL2002} found a correlation using a sample of 18 novae with more than one independent source for the mass of the ejecta of most systems. These results are shown in figure \ref{figEjectat2}. \citet{DellaValleETAL2002} provide the fitting relation (with a $95\%$ confidence level):
\begin{equation} \label{C2eqCorrel}
Log(M_{\text{ej}}) = 0.274(\pm 0.197) \cdot Log(t_2) + 0.645(\pm 0.283),
\end{equation}
with $M_{\text{ej}}$ measured in $10^{-5} $ M$_\odot$ and $t_2$ in days. 

Applying equation \eqref{C2eqCorrel} with $M_{\text{ej}} = 2 \cdot 10^{-6} $ M$_\odot$ gives a very short predicted decline time-scale: the higher value in the interval permitted by equation \eqref{C2eqCorrel} is $t_2 \lesssim 1$ d, while the observed decline time-scales can be estimated from the light curves as $t_2 \approx 20$ d for CI Aql and $t_2 \approx 30$ d for T Pyx. This places CI Aql and T Pyx significantly off the curve of equation \eqref{C2eqCorrel}. We conclude that the optical decline time-scales of these systems are significantly longer than those of systems with a similar value of $M_{\text{ej}}$.

\subsection{X-Rays emission} \label{sec:xrays}
Classical novae, including T Pyx, IM Nor, CI Aql, and other recurrent systems, show an extended strong soft X-rays emission after the outburst which is modeled as continuing nuclear reactions in the white dwarf envelope. The peak photon energy is $E_{\text{peak}} < 1$ keV (hence the often used term \emph{supersoft}), and the FUV/X-rays spectrum resembles a blackbody distribution only at low resolution (see \citealt{OrioETAL2001} and \citealt{Schwarz2011}). Since the ejecta of the system are also expelled from the white dwarf at the beginning of the eruption, a correlation is expected between the duration of the supersoft emission, the mass of the ejecta, and the decline time-scale of the system.

\begin{figure}
\begin{center}
\includegraphics[width=0.5\textwidth]{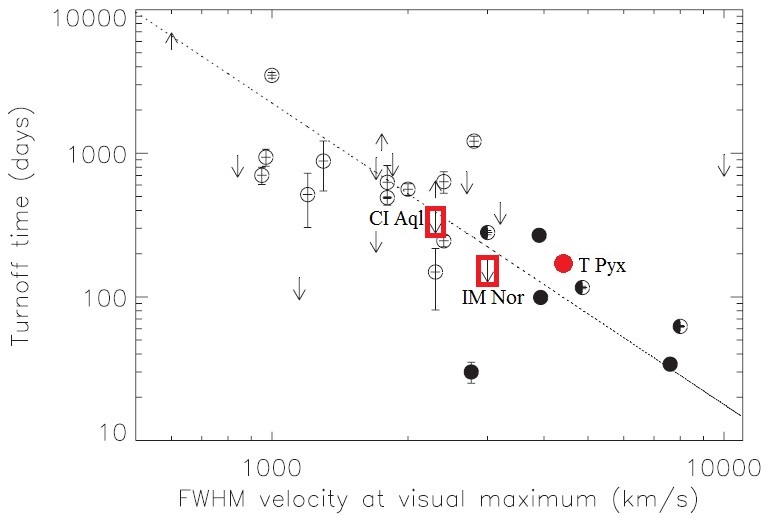}
\caption{\label{figXRaysVelocity}Correlation between the velocity of the ejecta and the duration of the soft X-rays phase. The x-axis shows the FWHM velocity of the $H\alpha$ and $H\beta$ spectral lines near visual maximum, an indication of the ejecta velocity. The y-axis gives the time at which the ratio of soft over hard emission drops below a certain threshold and is an indication of the turnoff time of the nuclear reactions at the white dwarf surface. Filled circles are known recurrent novae. Half filled circles are suspected recurrent novae based on their characteristics. The arrows indicate upper (pointing down) and lower (pointing up) boundaries. The dashed line shows the fit by \citet{Greiner2003} based on a subset of the reported data. Adapted from \citet{Schwarz2011}.}
\end{center}
\end{figure}

 \citet{Schwarz2011} performed a correlation study of the \emph{velocity} of the ejecta, which is related to their mass and can be directly obtained spectroscopically, and the duration of the X-rays emission.  We show their results in figure \ref{figXRaysVelocity} where we have marked the position of T Pyx (see section \ref{sec:TPyxXRays}), CI Aql, and IM Nor. In the sense of this correlation, the supersoft turn-off times of these three systems are not especially anomalous.

\begin{figure}
\begin{center}
\includegraphics[width=0.5\textwidth]{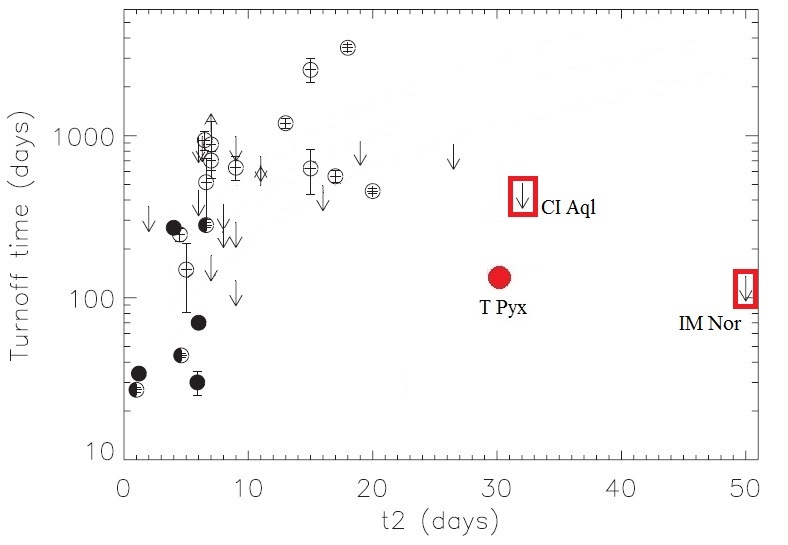}
\caption{\label{figXRayst2}Correlation between the optical decline time-scale $t_2$ and the turnoff time. As in figure \ref{figXRaysVelocity}, filled circles are known recurrent novae; half filled circles are suspected recurrent novae based on their characteristics; the arrows indicate upper (pointing down) and lower (pointing up) boundaries. The approximate position of T Pyx is depicted as a red circle, showing its peculiar behaviour. The upper boundaries for CI Aql and IM Nor, already included in the analysis by \citet{Schwarz2011}, have been highlighted. Figure adapted from \citet{Schwarz2011}.}
\end{center}
\end{figure}

\citet{Schwarz2011} have also studied the correlation between the decline time-scale $t_2$ and the turn off of the X-rays emission. Their results are shown in figure \ref{figXRayst2}, where we have again marked the position of T Pyx, CI Aql, and IM Nor. Although the scatter is significant and there is a scarcity of systems with long decline time-scales, some correlation is evident. T Pyx, IM Nor, and CI Aql are clearly deviant in this correlation.  We conclude from the analysis of figures \ref{figXRaysVelocity} and \ref{figXRayst2} that while the X-rays emission and ejecta velocities of these systems are similar to those of the other recurrent novae, their decline time-scales $t_2$ are anomalous.

\subsubsection{A word of caution: emission from T Pyx} \label{sec:TPyxXRays}
The X-rays emission from T Pyx has been studied with the \emph{Swift} and \emph{Suzaku} facilities. The results are discussed by \citet{DeGennaroETAL2014} and \citet{Chomiuk2014}. The X-rays behaviour of T Pyx is anomalous, including the time evolution of the hard-to-soft X-rays emission ratio which is used by \citet{Schwarz2011} to define the turn off time of the phase of soft X-rays emission. \citet{DeGennaroETAL2014} suggest that the emission received from T Pyx might have been produced by the hot ejecta and not by the central source, while \citet{Chomiuk2014} discuss the possibility of a dual-phase ejection, with a two-months interval between the ejections.

To include T Pyx in the correlation studies of the other novae, we adopted the turnoff time $t_{\text{off}} \approx 140$ d, again given by \citet{DeGennaroETAL2014}. The previous arguments would still hold for the other two systems if T Pyx were to be excluded for its peculiarity.

\section{Reprocessing of the radiation during the outburst} \label{sec:effect}
\begin{figure*}
\begin{center}
\includegraphics[width=\textwidth]{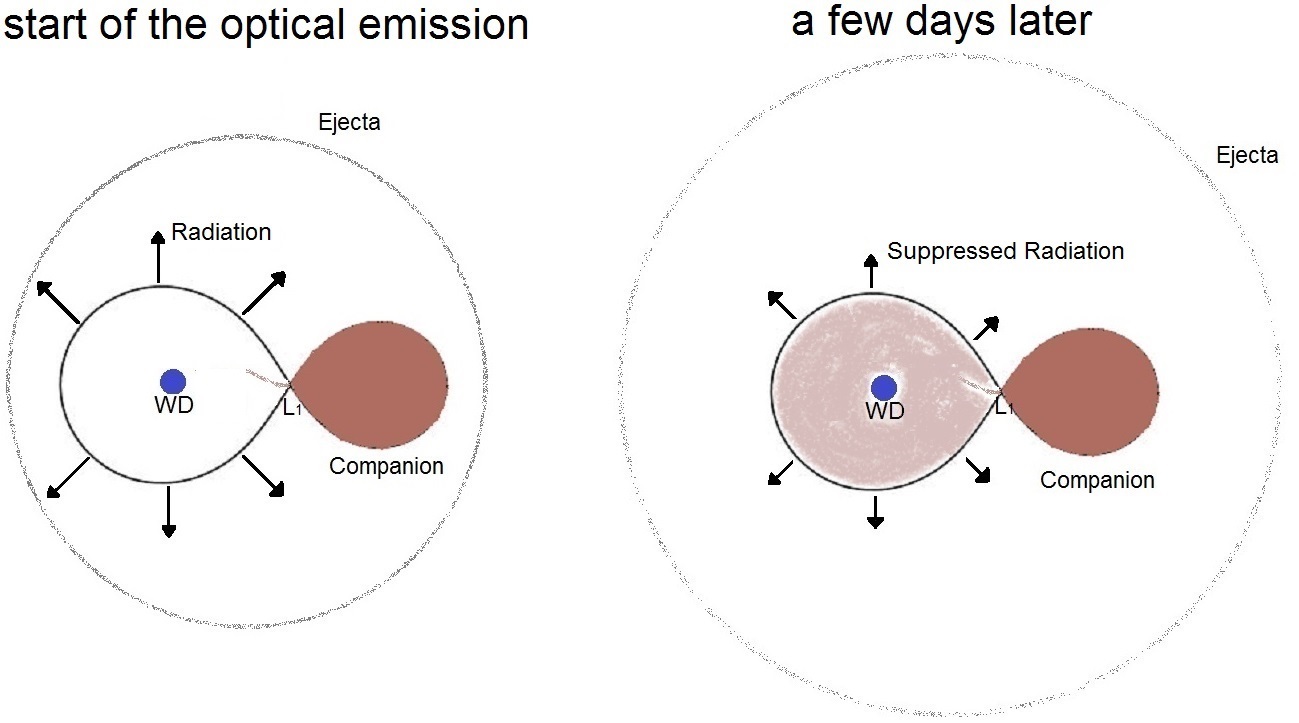}
\caption{\label{figScenario}Schematic illustration of the scenario discussed in section \ref{sec:effect} for T Pyx, IM Nor, and CI Aql. The optical emission starts at the end of the fireball phase. The material in the Roche lobe of the white dwarf has been blown away by the eruption, but the mass transfer from the companion through the $L_1$ point is uninterrupted. After a few days, the expansion of the ejecta has caused their density to drop. However, the material in the Roche lobe of the white dwarf suffices to absorb most of the radiation (illustrated by arrows in the figure) and re-emit it at a much lower temperature, decreasing its effectiveness. The ejecta are not ionized and the optical emission continues. The figure is not to scale: the radius of the ejecta a few days after the eruption in a scaled figure would be considerably larger.}
\end{center}
\end{figure*}

In light of these observations, we propose the following scenario for the peculiar behaviour of T Pyx, IM Nor, and CI Aql:

\begin{itemize}
\item{Mass transfer from the companion star onto the white dwarf resumes immediately after the eruption with transfer rate $\dot M$.}
\item{A fraction $f$ of the transferred material orbits in the Roche lobe of the white dwarf, with density distribution $\rho(\bb{r}, t)$.}
\item{In most systems, after a time $\Delta t$ that is possibly as short as a few days, radiation from the white dwarf ionizes the ejecta and the optical luminosity declines.}
\item{In a few systems, whose particular properties we will discuss, the accreted material of mass $f \dot M \Delta t$ absorbs the radiation through photoionization before it hits the ejecta, impeding their ionization. The energy is then reemitted at a much lower temperature, and is less effective in ionizing the ejecta.}
\end{itemize}
We illustrate this scenario schematically in figure \ref{figScenario}, which shows the state of the system at the start of the optical emission (on the left) and after a time $\Delta t$ (on the right). For convenience, we have not drawn the figure to scale: the radius of the ejecta a few days after the eruption in a scaled figure would be considerably larger.

This screening effect is favoured in systems with a high mass transfer rate, that implies higher density within the Roche lobe of the white dwarf. The mass transfer rates of T Pyx and CI Aql are known to be high. The conventionally accepted values are $\dot M_{\text{T Pyx}} = (1 \div 5) \cdot 10^{-8}$ M$_\odot$ yr$^{-1}$ \citep{Selvelli2008} and $\dot M_{\text{CI Aql}} \approx 10^{-7}$ M$_\odot$ yr$^{-1}$ \citep{Hachisu2003}. There are no estimates of $\dot M$ for IM Nor. A recent analysis by \citet{Godon2014}, based on the fit of disc models to the data of far ultraviolet spectroscopy, proposes even higher values of $\dot M_{\text{T Pyx}}$, of order $10^{-6}$ M$_\odot$ yr$^{-1}$.

We propose that this effect is also favoured for systems with a low orbital period for the following reason. A low orbital period is related to a smal separation, and a small volume of the Roche lobe of the white dwarf. Thus, for a given accumulated mass, the region around the white dwarf should be at higher density $\rho(\bb{r}, t)$. Since the effectiveness of the orbiting material in absorbing the radiation depends on the recombination rate of its ionized atoms, which is proportional to $\rho^2$, a high density is a key element for the occurrence of this screening effect.

It is very difficult to devise a comprehensive model of the irradiation of the ejecta taking into account the transfer of material into the Roche lobe of the white dwarf, because the hydrodynamics of the flow of matter during the peak of the irradiation is very complex. Such model is beyond the scope of our present paper. Instead, we present a schematic model for estimating the amount of material required for the shielding of the ejecta to occur.

\subsection{The model} \label{sec:model}
We assume that the mass transfer rate after the eruption is the same as in the quiescent state (see appendix B for discussion), and that the radiation from the white dwarf is time independent and has a blackbody distribution. We assume that material around the white dwarf has a spherically symmetric distribution $\rho(r)$ and uniform temperature $T_{\text{m}}$. Our aim is to determine the minimum value of the fraction of material $f$ that is needed to absorb the radiation incoming from the white dwarf after a time $\Delta t$ since the eruption. Any result $f > 1$ would imply that the effect we are describing is unrealistic.

There are few \emph{a priori} constraints for the value of $f$ from hydrodynamics. To our knowledge, there are no models of accretion disc formation during the peak of the irradiation of a nova. General simulations, as those by \citet{Lanzafame2006} (see their figure 2), still allow a great range of values for $f$, which varies from effectively 0 for high viscosity material to about $10^{-1}$ for very low viscosity values.

We adopt the following fixed numerical values for the system:

\begin{itemize}
\item{For the white dwarf, a mass of 1 M$_\odot$, and a radius of 0.01 R$_{\odot}$. For the secondary, a mass of 0.2 M$_\odot$ (these are reasonable but debatable values for the T Pyx system; see \citet{UthasArticle}).}
\item{Temperature of the white dwarf $T_{\text{WD}} = 7 \cdot 10^5$ K, corresponding to Eddington luminosity.}
\item{Time that would be required for the ionization of the ejecta without the intervening matter $\Delta t = 5$ d, which is a typical decline time-scale of very fast classical novae \citep{Strope}.}
\item{Temperature of the material in the Roche Lobe of the white dwarf: $T_{\text{m}} = 10^4$ K. This value is consistent with energy balance in the Roche lobe; see appendix C1 for discussion.}
\item{For the opacity, we neglect the neglect the metals in the chemical composition of the transferred material and adopt for the hydrogen and helium abundances: $X = 0.75$, $Y = 0.25$. This facilitates the calculations adopting a lower limit for the total opacity. See appendix C2 for a discussion.}
\end{itemize}

We label as $N_\nu (r)$ the number of photons per unit frequency at a distance $r$ from the white dwarf, and proceed to determine $N_\nu (r)$ as the number of photons emitted from the surface of the white dwarf minus those absorbed by the material in the Roche lobe. The photons are absorbed by processes of photoionization when they strike the atoms; we neglect the scattering processes because they are approximately elastic and have a smaller effect on $N_\nu (r)$.

At a temperature $T_{\text{m}} = 10^4$ K, part of the hydrogen is ionized because of the collisions and does not contribute to the absorption. The amount of ionized atoms depends on the density of the material. Our model takes this effect into account by considering a reduced number of H atoms, given by the fraction of atoms that are not collisionally ionized, as derived by solving the Saha equation. We label such fraction as $F(\rho)$. At $T_{\text{m}} = 10^4$ K, for the densities relevant to our scenario, only a negligible fraction of the He atoms are ionized by collisions. We acknowledge that our description constitutes an oversimplified view of the complex interaction between the radiation from the white dwarf, the collisions between the atoms and those between the atoms and the electrons, and the recombination processes, but we proceed in this way as our aim is to obtain an order-of-magnitude estimate of $f$ rather than an exact value. 

We label as $B_\nu(T_{\text{WD}})$ the black body spectrum at the white dwarf surface temperature, as $\sigma_i(\nu)$ the photoionization cross section for photons of frequency $\nu$ striking atoms of species $i$, as $\alpha_i(T)$ the case B recombination coefficient of the species $i$, as $n_{\text P}$, $n_\alpha$ and $n_{\text e}$ the number densities of H, He nuclei (i.e. protons and $\alpha$ particles) and electrons. In what follows, $\bar n_i(r)$ is the number density of the species $i$ that contribute to the recombination - photoionization balance, so that $\bar n_{\text P} = F(\rho) n_{\text P}$ and $\bar n_{\alpha} = n_{\alpha}$. We denote as $R_{\text{RL}}$ the radius of the Roche lobe and compute it by means of Eggleton's formula \citep{Eggleton}.

Analytical formulae for $\alpha_H$, $\alpha_{He}$ and for the cross section $\sigma_{\text{H} \rightarrow \text{H}^+}(\nu)$ are given in \citet{Draine}. The recombination coefficient $\alpha_{\text{He}^{++}}$ is derived from the hydrogen recombination coefficient as described in section 4.2 of \citet{Osterbrock}. The He photoionization cross section $\sigma_{\text{He} \rightarrow \text{He}^+}(\nu)$ is interpolated from the data by \citet{Yan}. The photoionization cross section of He$^{+}$ is given in section 5.4 of \citet{Allen}. 

The number densities of protons (either free or in H atoms) and $\alpha$ particles (either free or in He atoms) are given by:
\begin{equation}
n_{\text P}(r) = X \rho(r) \frac{1}{m_{\text{P}}},
\end{equation}
\begin{equation}
n_\alpha(r) = Y \rho(r) \frac{1}{m_\alpha},
\end{equation}
where $m_{\text{P}}$ and $m_\alpha$ are the masses of the H and He nuclei respectively.

The equations that give the number of photons radiated from the white dwarf and their absorption by the material in the Roche lobe are:
\begin{flalign}
& N_\nu (R_{\text{WD}}) = 4 \pi R_{\text{WD}}^2 B_\nu (T_{\text{WD}}) \frac{1}{h \nu}, & \\
& \frac{d N_\nu}{dr} = - N_\nu \Big( \sum_{i=\text{H,He,He}^+}{\sigma_i(\nu) \bar n_i (r)} \Big), &
\end{flalign}
The equations that give the balance between the recombination and the ionization of H, H$^+$, He, He$^+$ and He$^{++}$ are:
\begin{flalign}
& \bar n_{{\text H}^+}  n_{\text e}  \alpha_{\text{H}^+} (T_{\text{m}})= \frac{\bar n_{\text H}}{4 \pi r^2} \int_0^{+\infty}{N_\nu  \sigma_{\text{H} \rightarrow \text{H}^+}(\nu) d\nu}, & \\
& \bar n_{{\text{He}}^+}  n_{\text e}  \alpha_{\text{He}^+} (T_{\text{m}})= \frac{\bar n_{\text {He}}}{4 \pi r^2} \int_0^{+\infty}{N_\nu  \sigma_{\text{He} \rightarrow \text{He}^+}(\nu) d\nu}, & \\
& \bar n_{{\text{He}}^{++}}  n_{\text e}  \alpha_{\text{He}^{++}} (T_{\text{m}})= \frac{\bar n_{{\text{He}}^+}}{4 \pi r^2} \int_0^{+\infty}{N_\nu  \sigma_{\text{He}^+ \rightarrow \text{He}^{++}}(\nu) d\nu}, &
\end{flalign}
where the dependence of $\bar n_{{\text H}^+} = \bar n_{{\text H}^+}(r)$, $\bar n_{\text{He}^+}$, $\bar n_{{\text{He}}^{++}}$, $n_{\text e}$ and $N_{\nu}$ on $r$ has been omitted for readability.
  
Finally, the equations of conservation of the total number of H and He nuclei and electrons are:
\begin{flalign}
& \bar n_{{\text H}^+} (r) + \bar n_{\text H} (r) =  F(\rho) n_{\text P}(r), &  \\ 
& \bar n_{{\text{He}}^{++}} (r) + \bar n_{{\text{He}}^+} (r) + \bar n_{{\text{He}}} (r) = n_\alpha(r), & \\ 
& (1-F(\rho)) n_{\text P} + \bar n_{{\text H}^+} (r) + \bar n_{{\text{He}}^+} (r)  + 2 \bar n_{{\text{He}}^{++}}(r) = n_{\text e} (r). &
\end{flalign}

Solving equations (5) - (11) in the variables $N_\nu$, $\bar n_{\text{H}}$, $\bar n_{\text{H}^+}$ , $\bar n_{{\text{He}}}$, $\bar n_{\text{He}^+}$, $\bar n_{{\text{He}}^{++}}$, $n_{\text e}$  with the boundary condition (4), we determine $N_\nu(r)$. We thus obtain $\beta$, the energy that is free to escape the Roche lobe:
\begin{equation}
\beta = \Big( \int_0^{+\infty}{h \nu N_\nu(R_{\text{RL}}) d \nu} \Big) \Big/ \Big( \int_0^{+\infty}{h \nu N_\nu(R_{\text{WD}}) d \nu} \Big).
\end{equation}

\subsection{Results} \label{sec:results}
To solve equations (5) - (11), a density distribution law $\rho(r)$ is required, for a total mass in the Roche lobe $f \dot M \Delta t$. Since the recombination rate depends on the square of the density, the required value of $f$ will depend strongly on the density profile. Unfortunately, the distribution of the material depends on the hydrodynamics of the accretion process and is therefore difficult to predict in advance. 

We discuss the simplest case of uniform density distribution $\rho (r) = \bar \rho$. It is important to note that this is the least favourable choice since the effectiveness of recombination grows strongly with $\rho$ and the radiation is therefore more easily absorbed if there are regions of high density and regions of low density rather than a uniform profile. As an illustrative example, we also present results for the case $\rho(r) \propto r^{-3/2}$. The $r^{-3/2}$ profile has physical significance because it is the density distribution that would result in the case of steady, free-fall, spherical accretion onto the white dwarf, and represents an extreme case worth studying.

As stated in section \ref{sec:model}, we assumed $M_{\text{WD}} =$ 1 M$_\odot$,  $M_{\text{comp}}$ = 0.2 M$_\odot$, $T_{\text{WD}} = 7 \cdot 10^5$ K. We extracted the orbital periods from table \ref{table:RN} and set $\dot M_{\text{T Pyx}} = 3 \cdot 10^{-8}$ M$_\odot$ yr$^{-1}$, in the mid-point of the range of values from \citet{Selvelli2008} and $\dot M_{\text{CI Aql}} = 10^{-7}$ M$_\odot$ yr$^{-1}$ as given by \citet{Hachisu2003}.

We consider the material in the Roche lobe to be effective in absorbing the radiated photons if 90\% of the emitted energy is absorbed, i.e. if $\beta = 0.1$. We denote as $f_{\text{c}}$ the fraction of transferred material that has to orbit in the Roche lobe for this to occur and give values of $f_{\text{c}}$ for T Pyx and CI Aql in the uniform density and $\rho(r) \propto r^{-3/2}$ cases in table \ref{table:results1}. 

\begin{table}
\centering
\begin{tabular}{| c | c | c |}
\hline
	& uniform density & $\rho(r) \propto r^{-3/2}$  \\ \hline
T Pyx	& $f_{\text{c}} = 2.5 \cdot 10^{-2}$ & $f_{\text{c}} = 1.1 \cdot 10^{-2}$ \\ \hline
CI Aql	& $f_{\text{c}} = 6.0 \cdot 10^{-2}$ & $f_{\text{c}} = 2.6 \cdot 10^{-2}$ \\ \hline
\end{tabular}
\caption{Fraction $f_{\text{c}}$ of transferred material that needs to stay in the Roche lobe of the white dwarf of T Pyx and CI Aql in order to absorb 90\% of the radiated energy.}
\label{table:results1}
\end{table}

In all cases, a fraction $f_{\text{c}} < 0.1$ is sufficient to absorb the outgoing radiation. For $\rho(r) \propto r^{-3/2}$, $f_{\text{c}}$ is of order $10^{-2}$. If the density distribution differs greatly from the ones we studied with our model, even smaller values of $f$ might suffice. Thus, the transferred material seems to be sufficient to shield the ejecta. In light of these results, we argue that the scenario we described is realistic.

We show in table \ref{table:results2} the $f - \beta$ correspondence for some values around $f_{\text{c}}$ for both T Pyx and CI Aql, in the case of uniform density distribution. $\beta$ depends strongly on $f$ so that a moderate variation in the amount of absorbing material can have a large effect.

\begin{table}
\centering
\begin{tabular}{ |c|c||c|c| }
  \hline
  \multicolumn{2}{|c||}{T Pyx} &  \multicolumn{2}{c|}{CI Aql}  \\
  \hline
  f & $\beta$ & f & $\beta$  \\ \hline
$  1.0 \cdot 10^{-2}$ & 0.9 &$  4.0 \cdot 10^{-2}$ & 0.7 \\
$  2.0 \cdot 10^{-2}$ & 0.4 & $ 5.0 \cdot 10^{-2} $& 0.4 \\
$  2.5 \cdot 10^{-2}$ & 0.1 & $ 6.0 \cdot 10^{-2}$ & 0.1 \\
$  3.0 \cdot 10^{-2}$ & 0.0 &  $7.0 \cdot 10^{-2}$  & 0.0 \\
  \hline
\end{tabular}
\caption{Fraction $f$ of material in the Roche lobe and fraction of non-absorbed energy $\beta$ for T Pyx and CI Aql, in the case of uniform density distribution.}
\label{table:results2}
\end{table}
\begin{table}
\centering
\begin{tabular}{ |c|c| }
  \hline
  $P_{\text{orb}}$ (d) & $f_{\text{c}}$ \\ \hline
  $0.05 $  &$  1.5 \cdot 10^{-2}$ \\
  $0.1   $  &$  3.4 \cdot 10^{-2}$ \\
  $0.2   $  &$  7.6 \cdot 10^{-2}$ \\
  $0.3   $  &$  1.2 \cdot 10^{-1}$ \\
  $0.5   $  &$  1.6 \cdot 10^{-1}$ \\
  $1      $  &$  3.3 \cdot 10^{-1}$ \\
  $2      $  &$  6.7 \cdot 10^{-1}$ \\
  $3      $  &$  1.0  $ \\
      
  \hline
\end{tabular}
\caption{Fraction $f_{\text{c}}$ of material corresponding to $\beta = 0.1$ for the mass transfer rate of T Pyx, as a function of the orbital period, in the case of uniform density distribution.}
\label{table:results3}
\end{table}

We argued that a longer orbital period $P_{\text{orb}}$ should lead to higher value of $f_{\text{c}}$. We show some values of $P_{\text{orb}} - f_{\text{c}}$ in table \ref{table:results3}, assuming a mass transfer rate $\dot M = 3 \cdot 10^{-8}$ M$_\odot$ yr$^{-1}$, as for T Pyx, and uniform density. As we expected, $f_{\text{c}}$ grows with $P_{\text{orb}}$. For systems with an orbital period of several days, not even all of the transferred material would suffice to shield the ejecta.

The results for $\beta$ depend uniquely on the product $f \dot M$, not on the individual values of $f$ and $\dot M$. For the values $\dot M_{\text{T Pyx}} \gtrsim 10^{-6}$ M$_\odot$ yr$^{-1}$ proposed by \citet{Godon2014}, lower values of $f$ are required to shield the ejecta, with $f_c$ of order $10^{-3}$ for the uniform density model and $10^{-4}$ for the $\rho \propto r^{-3/2}$ model.

\section{Conclusion} \label{sec:conclusion}
In this paper, we discussed some properties of the three recurrent novae with the lowest orbital periods, T Pyx, IM Nor, and CI Aql, noting that they have the longest optical decline time-scales but the mass and velocity of their ejecta and the duration of their X-rays emission are similar to those of systems with more rapid optical declines.

We put forth a scenario to explain the reason for this occurrence. We propose that some of the material transferred from the secondary star during the eruption contributes to the absorption of the radiation from the white dwarf and shields the ejecta before their ionization occurs. Quantitative evaluation of this scenario indicates that for a system with a large mass transfer rate and short orbital period, the amount of transferred material is indeed sufficient. A more thorough analysis of this problem requires understanding the hydrodynamic of the transfer process. Unfortunately the situation is so complex, not to mention the presence of the ejecta and the irradiation, that the feasibility of any model on this subject is questionable. 

The detailed physics of the ejecta has not been addressed in this study, but it plays a key role in determining the photometric properties of the system. While it is clear that the radiation reprocessed from the material in the Roche lobe of the white dwarf is reemitted at a much lower temperature, and is therefore less effective in ionizing the ejecta, its effect on the ejecta should be studied to gain a deeper understanding of the light curve of the system. 

Another key aspect that remains to be investigated is the consequences of the shielding effect on the spectrum of the system. Although the spectrum of T Pyx during various phases of its outburst has been observed (see \citealt{ShoreETAL2012}, \citealt{ShoreETAL2013}), the major obstacle in understanding the behaviour of this particular system is the complexity of the spectra rather than the scarcity of the data. The analysis of the spectra prompted \citet{ShoreETAL2013} to suggest that T Pyx is peculiar with respect to most of the classical novae and that its \emph{`extended opaque phase could be due to the formation of a cool common envelope after the explosion, causing a recombination wave to move outward through the ejecta that also extinguishes the XR emission. It could then slowly clear as the WD settles into a stage of quasi-static nuclear burning and develops a supersoft source'}. The scenario we described provides an explanation for how such an envelope could be formed and motivates further thought on the spectra of the system.

\section*{Acknowledgements}
Andrea Caleo acknowledges support from the University of Oxford.  We thank Ivan De Gennaro Aquino, Jordi Jos\'e, Elena Mason, Kim Page and Ashley Wagner for many useful discussions and comments.

\appendix

\section{Orbital period of T Pyx before and after the eruption}
\citet{Patterson2014} reported a campaign to track the photometric wave of T Pyx and follow the evolution of its orbital period, gathering data over the 1996-2011 period. They reported a gradual increase of the orbital period during the quiescent phase and a jump of $+0.0054(6)\%$ at the eruption. They claim that this implies that the mass of the ejecta of T Pyx is at least $3 \cdot 10^{-5}$ M$_{\odot}$. We argue here that the variation of the orbital period doesn't imply a lower bound to the ejected mass.

\citet{Patterson2014} state that \emph{`during the eruption, mass loss should increase $P_\text{orb}$, and angular-momentum loss should decrease it'}. A positive change in the period would imply that the mass loss effect wins and the mass loss is at least $3 \cdot 10^{-5}$ M$_\odot$. Although no derivation is shown, this result appears to be based on the analysis by \citet{Livio1991}, which in turn is based on that by \citet{Shara1986}. However, a tacit but fundamental assumption is at the core of the paper by Shara et al.: the formulas reported, most notably their equation (5), are only valid if the orbit of the system before and after the eruption is circular. While it is possible to imagine a gradual circularization of the orbit after the eruption, so that this assumption might hold for the study of the long-term evolution of the system, the eccentricity right after the eruption is likely non-zero.

In general, the variation of the period of a system in a circular or elliptical orbit depends on the change of the masses of its components and the mechanical energy of the system. Equation (11) of \citet{SaasFee3} relates the change in the period $P$ of a binary system to the change of its total mass $M$ and its energy per unit reduced mass $\epsilon$:
\begin{equation}
\frac{dP}{P} = \frac{dM}{M} - \frac{3}{2} \frac{d \epsilon}{\epsilon}
\end{equation}
We express this result in terms of the masses $m_1$, $m_2$ of the components of the system and in terms of the total energy $E = \mu \epsilon$ (where $\mu = (1/m_1 + 1/m_2)^{-1}$ is the reduced mass):
\begin{equation}
\frac{dP}{P} = \frac{dm_1}{M} \Big( 1 + \frac{3}{2} \frac{m_2}{m_1} \Big) + \frac{d m_2}{M} \Big( 1 + \frac{3}{2} \frac{m_1}{m_2} \Big) - \frac{3}{2} \frac{dE}{E}
\end{equation}
For mass loss from the primary star, $d m_1 = -M_{\text{ej}}$, $d m_2 = 0$:
\begin{equation} \label{PattersonCorrect}
\frac{\delta P}{P} = - \frac{M_\text{ej}}{M} \Big( 1 + \frac{3}{2} \frac{m_2}{m_1} \Big) - \frac{3}{2} \frac{\delta E}{E}
\end{equation}
where $\delta P$ is the period variation and $\delta E$ is the change of mechanical energy of the system. It is very complex to even estimate $\delta E$ from models of the ejection mechanism (during which nuclear energy is converted to thermal energy through degenerate H burning at the surface of the white dwarf, and then to mechanical energy of the ejecta and possibly the binary system); even the kinetic energy of the ejecta is currently not very well known (see e.g. the discussion in \citealt{Shara2010}). Without an estimate of $\delta E$, equation \eqref{PattersonCorrect} cannot provide an estimate of $M_\text{ej}$.

\section{Mass transfer during the eruption}
In the analysis of section \ref{sec:effect}, the mass transfer rate from the companion to the white dwarf during the first few days of the eruption was considered to be the same as that at quiescence. We discuss here this assumption. 

The secondary star is strongly irradiated during the outburst. It could be argued that this would cause an expansion of the star and an increase in the mass transfer rate. The problem of irradiation-induced mass transfer in novae has been widely discussed in the literature; however, the focus has always been on the long term evolution of the system rather than the first few days after the outburst. The most influential work in the field was conducted by \citet{KovetzETAL1988}. These authors determined the depth reached by the irradiation and argue that the main scattering process in the atmosphere of the star is electron scattering. They obtained for the mass of the irradiated layer $M_{\text{irr}} \approx 5 \cdot 10^{-8} M$, where $M$ is the mass of the companion. The companion would then inflate on a time-scale of $\approx 0.1$ yr, overfilling its Roche lobe and causing a significant increase in the mass transfer rate.

In contrast, we are considering the evolution of the system on a very short time-scale, the first few days after the outburst. Since the star would not have had time to fully expand, the scenario proposed by \citet{KovetzETAL1988} does not apply to our case. Moreover, we argue that the incoming photons are absorbed and re-emitted by processes of photoionization rather than electron scattering and give an independent estimate of $M_{\text{irr}}$.

We achieve an order-of-magnitude estimate of $M_{\text{irr}}$ by assuming that the atmosphere of the companion is composed of hydrogen, with a density profile similar to that of a non-Roche-lobe-filling main sequence star of the same mass (we acknowledge that this is inaccurate in a neighbourhood of the $L_1$ point). We make use of the density profile of a star with mass $M \approx 0.2$ M$_\odot$, effective temperature $T_{\text{eff}} = 3200$ K, gravity $Log(g) = 5$, and solar chemical composition extracted from the NextGen models  (see \citet{NextGen}). We evaluate the effect of a monochromatic irradiation at the peak wavelength of the emission from the white dwarf, at $T_{\text{WD}} = 7 \cdot 10^5$ K as in section \ref{sec:model}. The resulting peak photon energy is $E_{\text{peak}} \approx 60$ eV.

Proceeding as in section \ref{sec:model}, we label as $N_\gamma(z)$ the number of photons per unit area and unit time that reach depth $z$ in the atmosphere, as $\bar n_{\text H}(z)$ and $\bar n_{{\text H}^+}(z)$ the number densities of H and H$^+$ that contribute to the recombination - photoionization balance, as $\sigma_{\text{ph}}$ the photoionization cross section of H atoms at $E_{\text{peak}} \approx 60$ eV, as $F(\rho)$ the fraction of H atoms that would not be collisionally ionized at density $\rho$ and temperature $T_{\text{m}}$, and as $\alpha(T_{\text{m}})$ the type B recombination coefficient of hydrogen. As in the case of section \ref{sec:model}, we assume $T_{\text{m}} = 10^4 K$; see appendix C1 for discussion. We label as $a$ the distance between the white dwarf and the secondary star. The equations that determine $N_\gamma (z)$ are:
\begin{flalign}
& N_\gamma(0) = \frac{L_{\text{WD}}}{4 \pi a^2} \frac{1}{E_{\text{peak}}}, & \\
& \frac{d N_\gamma}{dz} = - N_\gamma(z) \bar n_{\text H}(z) \sigma_{\text{ph}}, &\\
& \bar n_{{\text H}^+}^2(z) \alpha(T_{\text{m}}) = \bar n_{\text H}(z) \sigma_{\text{ph}} N_\gamma(z), &\\
& \bar n_{\text H}(z) + \bar n_{{\text H}^+}(z) = \bar n_{\text P}(z), & \\
& \bar n_{\text P}(z) = F(\rho_{\text{NextGen}}(z)) \frac{\rho_{\text{NextGen}}(z)}{m_{\text{P}}}. &
\end{flalign}
We solve equations (B2) - (B5) with the boundary condition (B1), and define the penetration depth $z_{\text{irr}}$ as the depth at which 90\% of the incoming energy has been absorbed. The result is $z_{\text{irr}} \approx 10^{7}$ cm. 

Assuming uniform irradiation of the exposed of the secondary star, we estimate $M_{\text{irr}}$ as:
\begin{equation} 
M_{\text{irr}} = 2 \pi R_2^2 \int_0^{z_{\text{irr}}}{\rho(z) dz} \approx 10^{-14} \text{M}_{\odot}.
\end{equation}
This result is six orders of magnitude lower than that of \citet{KovetzETAL1988}. The effect of the irradiation on the secondary star is therefore very small and the mass transfer rate is unlikely to be affected by it.

\section{Assumptions in the model of section 3.1}
\subsection{Temperature in the Roche lobe of the white dwarf}
In section \ref{sec:model}, we argued that the temperature $T_{\text{m}}$ of the material in the Roche lobe of the white dwarf is $T_{\text{m}} \approx 10^4$ K. We discuss here this assumption.

We use an energy argument to determine an approximate value for $T_{\text{m}}$. The material in the Roche lobe is heated because of the radiation incoming from the white dwarf. It dissipates this energy by processes of photo-recombination to excited states that release photons that are less energetic than the ones from the white dwarf. The mean free path of these photons in the Roche lobe is:
\begin{equation}
l_{\text{mfp}} = \frac{1}{\rho \kappa_{\text{bf}}(\rho, T_{\text{m}})}.
\end{equation}
We use Kramer's formula for the bound-free opacity $\kappa_{\text{bf}}$, assuming metallicity $Z = 0.02$ and H abundance $X = 0.75$. For the case of uniform density in the Roche lobe of T Pyx described in section \ref{sec:results}, with $f= f_{\text{c}} = 2.5 \cdot 10^{-2}$, this gives $l_{\text{mfp}} = 3 \cdot 10^9$ cm. The radius of the Roche lobe is of order $r_{\text{RL}} \sim 10^{10}$ cm. This means that, depending on the position in the Roche lobe, a significant fraction of the re-emitted photons are not re-absorbed and contribute to the cooling of the system; most of them, however, are not lost, so that our scenario is only approximately correct. However, $\kappa_{\text{bf}} (\rho, T_{\text{m}}) \propto T_{\text{m}}^{-3.5}$ depends strongly on $T_{\text{m}}$, so that, for a temperature $T = 1.2 \cdot 10^4$ K, $l_{\text{mfp}}$ and $r_{\text{RL}}$ are of the same order. To obtain a temperature estimate, we assume transparency for photons from recombinations and that these are the main source of cooling.

The energy per unit time and volume absorbed from the incoming radiation is:
\begin{equation} 
\varepsilon_{\text{abs}}(r) =  \frac{1}{4 \pi r^2} \frac{d}{dr} \int_0^{+\infty}{N_\nu h \nu d\nu}.
\end{equation} 

The energy per unit time and volume emitted by the material is:
\begin{equation} 
\varepsilon_{\text{em}} = \Lambda(T) n_{\text e} n_{{\text H}^+},
\end{equation} 
where $\Lambda(T)$ is the cooling function. We use the values tabulated by \citet{Sutherland} for solar composition (i.e. $[Fe/H] = 0$). Such cooling function was computed under conditions valid for the interstellar medium rather than a stellar environment with considerably higher density (of order $\rho_{\text{RL}} \sim 10^{-10}$ g cm$^{-3}$ in the Roche lobe of the white dwarf); however, in both cases the main contribution to the cooling function at $T \gtrsim 10^4$ K is from hydrogen recombination to excited states (see e.g. \citet{Dalgarno}), so that the predicted cooling rate is still approximately correct. Figure 8 of \citet{Sutherland} shows the cooling function for a large range of temperatures and various chemical compositions.

\begin{figure}
\begin{center}
\includegraphics[width=0.5\textwidth]{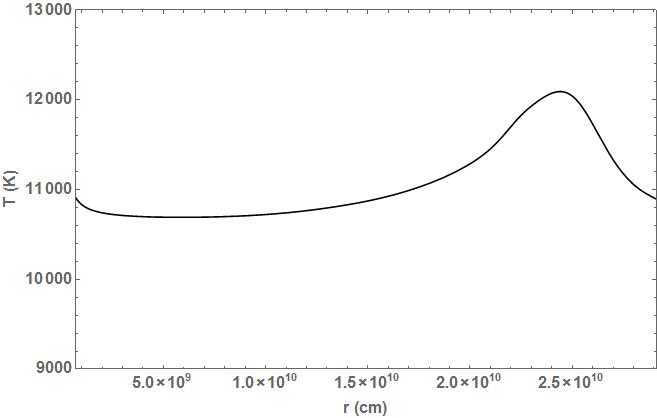}
\caption{\label{figTemperature}Temperature profile obtained imposing $\varepsilon_{\text{em}}(r, T) = \varepsilon_{\text{abs}}(r)$ for the case of uniform density for T Pyx with $f = f_{\text{c}}$.}
\end{center}
\end{figure}

We solve the equation $\varepsilon_{\text{em}}(r, T) = \varepsilon_{\text{abs}}(r)$ in the variable $T$ for every $r$ with the values of $\frac{d N_\nu(r)}{dr}$, $n_{\text H}^+(r)$, $n_{\text e}(r)$ for the uniform density case for T Pyx with $f= f_{\text{c}}$. This temperature profile is not self-consistent, as $\varepsilon_{\text{abs}}$ has been obtained by assuming a uniform $T_{\text{m}} = 10^4$ K profile, and will not reproduce in details the real profile in the Roche lobe. The temperature maximum in figure \ref{figTemperature} is located where the bulk of the radiation is absorbed from the material in the Roche lobe; here a significant number of atoms are neutral and the cooling is most efficient.

Our results show that the temperature in the Roche lobe, though not uniform, does not deviate significantly from $10^4$ K. This is due to the fact that the cooling function $\Lambda(T)$ varies very steeply with $T$ for $T \gtrsim 10^4$ K (see again figure 8 of \citealt{Sutherland}): a temperature profile that spans a range of a few $10^3$ K is suited to describe regions with significantly different rates of emitted energy. On the other hand, the recombination coefficients $\alpha_i$ and the fraction of non collisionally ionized H atoms $F(\rho)$ do not depend on $T$ in a similar manner, and they can be safely computed at the uniform temperature $T_{\text{m}} = 10^4$ K. This is why we have not considered the effects of a temperature gradient in the Roche lobe. Solving equations (5) - (11) of section \ref{sec:model} with the higher value $T = 1.2 \cdot 10^4$ K gives very similar results for the value of $f_{\text{c}}$ and for the temperature profile of figure \ref{figTemperature}.

Although we have focused the discussion of this appendix on just the case of section \ref{sec:effect}, a similar argument holds for the assumption $T_{\text{m}} = 10^4$ K of appendix B. In that case, almost all of the photons originated in recombination to excited states contribute to the cooling of the system, because the depth $z_{\text{irr}} \sim 10^{7}$ cm reached by the irradiation is at low optical depth, as can be seen from the data in the NextGen model ($\tau \lesssim 10^{-4}$). We conclude that for both problems, $T_{\text{m}} \sim 10^4$ K is an acceptable approximation and the temperature gradient is not very significant.

\subsection{Metals in the Roche lobe of the white dwarf}
In section \ref{sec:model}, we neglected the absorption of photons originated at the surface of the white dwarf by elements heavier than He. We discuss here this assumption.

Inspection of the solutions of equations (5) - (11) in the $\beta=0.1$ cases described in section \ref{sec:results} shows that, in all of these cases, most of the atoms, whether of H or He, are ionized. In this situation, the number of photons per unit volume and time absorbed for effect of the species $X$, equal to the number of recombinations that take place, is approximated by:
\begin{equation} 
\frac{d N_\gamma}{dt} \Big|_X = n_{X^+} n_\text{e} \alpha_X,
\end{equation} 
where $\alpha_X$ is the recombination coefficient of the species $X$. The relevance of the species $X$ to the absorption depends therefore on the value of the product $n_{X^+} \alpha_X$.

Assuming a metallicity of order $10^{-2}$, the total number density of the heavy elements with atomic number $Z \geq 6$ is of order $10^{-3}$ times that of the H atoms. Some recombination coefficients $\alpha_X$ have been tabulated, including those of the most common hydrogenic atoms \citep{Storey1995}. They are generally higher than the H recombination coefficient $\alpha_{\text{H}}$, but most often well within a factor of $10^2$ from it. We conclude that the contribution of the heavy elements to the absorption is of order $10^{-1}$, or lower, compared to that of the H atoms. It is therefore possible to neglect it in a schematic model as the one of section \ref{sec:model}.

\bibliographystyle{mn2e}
\bibliography{References}
\bibdata{References}

\label{lastpage}

\end{document}